\documentclass[structabstract]{aa}

\usepackage{graphicx}
\usepackage{txfonts}
\usepackage{natbib}
\bibpunct{(}{)}{;}{a}{}{,} 
\usepackage{amssymb}

\newcommand{\funits}{~$\rm erg~cm^{-2}~s^{-1}$}
\newcommand{\frequnits}{~$\rm s^{-1}$}

\newcommand{\lengthunits}{~$R_\odot$}
\newcommand{\velunits}{~$\rm km~s^{-1}$}

\newcommand{\specangunits}{~$\rm dyne~cm~sr^{-1}$}
\newcommand{\noflxunits}{~$\rm cm~s^{-2}$}

\usepackage{color}

\newcommand{\green}{}

\begin{document}
\title{Angular momentum transport in a multicomponent solar wind with differentially flowing, thermally anisotropic ions}
\author{Bo Li 
\and Xing Li}
\titlerunning{Angular momentum transport in multicomponent winds}
\authorrunning{Bo Li \& Xing Li}
\institute{Institute of Mathematics and Physics, Aberystwyth University,
  SY23 3BZ, UK. \\
  \email{[bbl, xxl]@aber.ac.uk}}

\date{Received / Accepted}

\abstract 
{
The Helios measurements of the angular momentum flux $L$ of the \green{fast} solar wind \green{lead to a tendency} for the fluxes associated with individual ion angular momenta of protons and alpha particles, $L_p$ and $L_\alpha$, to be negative (i.e., in the sense of counter-rotation with the Sun). However, \green{the opposite holds for the slow wind}, and the overall particle contribution $L_P = L_p+L_\alpha$ tends to exceed the magnetic
  contribution $L_M$. These two aspects are at variance with previous models.}
 {We examine whether introducing realistic ion temperature anisotropies can resolve \green{this} discrepancy.} 
 {From a general set of multifluid transport equations with gyrotropic species pressure tensors, we derive the equations governing both the meridional and azimuthal dynamics of outflows from magnetized, rotating stars. The equations are not restricted to radial flows in the equatorial plane but valid for general axisymmetric winds that include two major ion species. The azimuthal dynamics \green{are} examined in detail, using the empirical meridional flow profiles for the solar wind, constructed mainly according to measurements made in situ.} 
 {The angular momentum flux $L$ is determined by the requirement that the solution to the total angular momentum conservation law is unique and smooth in the vicinity of the Alfv\'en point, defined as where the combined Alfv\'enic Mach number $M_T=1$. $M_T$ has to consider the contributions from both protons and alpha particles. Introducing realistic ion temperature anisotropies may introduce a change of up to $10\%$ in $L$ and up to $\sim 1.8$\velunits\ in azimuthal speeds of individual ions between 0.3 and 1~AU, compared with the isotropic case. The latter has \green{strong} consequences on the relative importance of $L_P$ and $L_M$ in the angular momentum budget.} 
{However, introducing ion temperature anisotropies cannot resolve the discrepancy between in situ measurements and model computations.
For the \green{fast-wind} solutions, while in extreme cases $L_P$ may become negative,  \green{$L_p$ never does}.
On the other hand, for the slow solar wind solutions examined, $L_P$ never exceeds $L_M$\green{,} even though $L_M$ may be \green{less} than the individual ion contribution\green{,} since \green{$L_p$ and $L_\alpha$ always have opposite signs for the slow and fast wind alike}.
 }

\maketitle

\section{Introduction}
The angular momentum loss of a rotating star due to its outflow influences 
     the rotational evolution of the star \green{considerably}, and is therefore of astrophysical significance
     in general~\citep[see e.g.,][]{WD_67, BelcherMacGregor_76, MestelSpruit_87, Bouvier_etal_97}.
However, direct tests of in situ measurements
     against theories such as \green{those} presented by \citet{WD_67}
     are \green{only possible} for the present Sun.
A substantial number of studies have been conducted and were compiled in the comprehensive paper 
     by \citet{Pizzo_etal_83}, who themselves paid special attention to the Helios measurements of specific
     angular momentum fluxes.
The measurements, further analyzed by \citet{MarschRichter_84a}, 
     are unique in that they allow 
     the individual ion contribution from protons $L_p$ and alpha particles $L_\alpha$
     to the solar angular-momentum loss rate {per steradian} $L$ \green{to be examined}.
For instance, despite the significant scatter, the data 
      exhibit a distinct trend for ${L}_p$ to be positive (negative) for solar winds
      with proton speeds $v_p$ below (above) 400\velunits.
A similar trend for ${L}_\alpha$ is also found on average.
The magnetic contribution ${L}_M$, on the other hand, is remarkably constant.
A mean value of ${L}_M = 1.6 \times 10^{29}$\specangunits\ can be
      quoted for the solar winds of all flow speeds
      and throughout the region from 0.3 to 1~AU.
For comparison, the mean values of angular momentum fluxes carried by ion flows
      in the slow solar wind are
      ${L}_p = 19.6$ and ${L}_\alpha = 1.3$~$\times 10^{29}$\specangunits~\green{\citep[see Table~II of][]{Pizzo_etal_83}}.
The overall particle contribution to $L$ is then 
      ${L}_P = {L}_p + {L}_\alpha= 20.9 \times 10^{29}$\specangunits,
      which tends to be larger than ${L}_M$.
It is noteworthy that a more recent study by \citet{Scherer_etal_01} showed how examining the \green{long-term}
     variation of the non-radial components of the solar wind velocity and the corresponding 
     angular momentum fluxes
     can help us \green{understand the heliospheric magnetic field better}.

Alpha particles should be placed on the same footing as protons from the perspective of solar wind \green{modeling}, given
      their non-negligible abundance and the fact that there tends to exist a substantial differential
      speed $v_{\alpha p}\equiv |\vec{v}_{\alpha p}|\mbox{sign}(|\vec{v}_\alpha | -|\vec{v}_p|)$.
As shown by the Helios measurements, a $v_{\alpha p}$ amounting to up to $20\%-30\%$ of the local proton speed
      may occur in both the fast 
      and slow solar \green{winds}~\citep{Marsch_etal_82a, Marsch_etal_82b}, with the latter being exemplified by an event \green{that} took place on day 117 of 1978, 
      \green{when} a positive $v_{\alpha p}\sim 100$\velunits\ was found at 0.3~AU~\citep{Marsch_etal_81}.
\green{That} on the average $v_{\alpha p}\approx 0$ in the slow wind simply reflects that 
      the events with positive and negative $v_{\alpha p}$ occur with nearly equal frequency~\citep{Marsch_etal_82a}.
As for the alpha abundance relative to protons, a value of $4.6\%$ ($0.4\%-10\%$) is \green{well-established} for the fast (slow)
      solar wind~\cite[e.g.,][]{McComas_etal_00}.
Therefore alpha particles can play an important role as far as the energy
      and linear momentum balance of the solar wind are concerned.
When it comes to the problem of angular momentum transport, it was shown that in interplanetary space
      not only the angular momentum flux carried by the alpha particles $L_\alpha$ but also that convected by the protons $L_p$
      are determined by the terms associated with $v_{\alpha p}$~\citep{LiLi06}.
This essentially derives from the requirement that the proton-alpha velocity difference vector be aligned with the instantaneous
      magnetic field.
As a consequence, these terms have no contribution to the overall angular momentum flux convected by the ion flow $L_P$, which
      turns out to be smaller than $L_M$ in all the models examined in the parameter study by \citet{Li_etal_07}.
This, together with the fact that $L_p$ is always positive (i.e., in the sense of corotation with the Sun), is at variance
      with the Helios measurements. 

A possible means to reconcile the measurements and the model computation is to incorporate the species temperature anisotropies.
This is because the total pressure tensor $\tens{P}=\sum_s \tens{p}_s$ summed over all species $s$ participates in
      the problem of angular momentum transport via the component $P^\Delta =P^\parallel-P^\perp$ where
      $\parallel$ and $\perp$ are relative to the magnetic field $\vec{B}$~\citep[see e.g.,][hereafter referred to as W70]{Weber_70}.
While the overall loss rate per steradian $L$ may not be significantly altered, the azimuthal speed of the solar wind and therefore
      the particle part of $L$ may be when compared with the isotropic case.
Note that in the treatment of W70 the solar wind was seen as a bulk flow and the ion species are not distinguished.
On the other hand, the formulation by \citet{LiLi06} did not take into account the pressure anisotropy, which is a salient feature
      of the velocity distribution functions for both protons and alpha particles as revealed by the Helios measurements~\citep{Marsch_etal_82a,Marsch_etal_82b}.
It therefore remains to be seen how introducing the pressure anisotropy influences individual ion azimuthal speeds.
Moreover, the simple, prescribed functional form for $P^\Delta$ assumed in W70 
      needs to be updated in light of the more recent particle measurements.

The aim of the present paper is to extend the W70 study in three ways.
First, we shall follow a multicomponent approach and examine the angular momentum transport in a solar wind comprising protons,
      alpha particles and electrons where a substantial proton-alpha particle velocity difference exists.
Second, although following W70 we use a prescribed form of $P^\Delta$ for simplicity, this prescription is based on
      the Helios measurements, and also takes into account other in situ and remote sensing measurements. 
Third, unlike W70 where the model equations are restricted to the equatorial plane, the equation set we shall derive
      is appropriate for a rather general axisymmetrical, time-independent, multicomponent, thermally anisotropic flow 
      emanating from a magnetized rotating star.
We note that a similar set of equations, which was also restricted to radial flows, was derived by \citet{I84} who worked
      in the corotating frame of reference and neglected the azimuthal dynamics altogether.
The functional dependence on the radial distance and flow speed of the magnetic spiral angle was prescribed instead.
His approach is certainly justifiable for the present Sun, but a
     \green{self-consistent treatment of the azimuthal dynamics is required when flows from other stars are examined.
This is because many stars either have a stronger magnetic field
      or rotate substantially faster than the Sun.}

The paper is organized as follows.
We start with section~\ref{sec_mathform} where a description is given for the general multifluid, gyrotropic transport
      equations, based on which the azimuthal dynamics of the multicomponent solar wind 
      is examined.
Then section~\ref{sec_presflow} describes the adopted meridional magnetic field and flow profiles.
The numerical solutions to the angular momentum conservation law are given in section~\ref{sec_numres}.
In section~\ref{sec_discussion}, we shall discuss how examining the angular momentum transport in a multicomponent solar wind
      can also shed some light on the spectra of ion velocity fluctuations induced by Alfv\'enic activities.
Finally, section~\ref{sec_summary} summarizes the results.
The equations of and a discussion on the poloidal dynamics are presented in the appendix.

\section{Mathematical formulation}
\label{sec_mathform}

Presented in this section is the mathematical development of
     the equations that 
     govern the angular momentum transport in a time-independent solar wind
     which consists of electrons ($e$), protons
     ($p$) and alpha particles ($\alpha$).
Each species $s$ ($s = e, p, \alpha$) is characterized by its mass $m_s$,
     electric charge $e_s$, number density
     $n_s$, mass density $\rho_s = n_s m_s$, velocity $\vec{v}_s$,
     and partial pressure tensor $\tens{p}_s$.
If measured in units of the electron charge $e$, $e_s$ may be expressed by
     $e_s = Z_s e$ with $Z_e \equiv -1$ by definition.

To simplify the mathematical treatment, a number of assumptions have been made
     and are collected as follows:
\begin{enumerate}
\item 
Symmetry about the magnetic axis is assumed, i.e.,
     $\partial/\partial\phi\equiv 0$ in a heliocentric spherical coordinate system
     ($r, \theta, \phi$).
\item
The velocity distribution function (VDF) of each species is close to a bi-Maxwellian, and
     the pressure tensor is gyrotropic, i.e., 
     $\tens{p}_s = p_s^{\perp} \tens{I} + (p_s^{\parallel}-p_s^{\perp})\hat{b}\hat{b}$, where
     $\tens{I}$ is the unit dyad and $\hat{b}$ is the unit vector along the magnetic field $\vec{B}$.
The temperatures pertaining to the degrees of freedom parallel and perpendicular to $\vec{B}$
     follow from the relation  
     $p_s^{\parallel, \perp} = n_s k_B T_s^{\parallel, \perp}$, 
     where $k_B$ is the Boltzmann constant.
\item 
Quasi-neutrality is assumed, 
      i.e., $n_e = \sum_k Z_k n_k$ .
\item 
Quasi-zero current is assumed, 
      i.e., $\vec{v}_e = \sum_k Z_k n_k \vec{v}_k/n_e$ ($k=p, \alpha$),
      except when the reduced meridional momentum equation is derived.
\end{enumerate}

\subsection{Multi-fluid equations}
\label{sec_mathform_multifluideqs}

The equations appropriate for a multi-component solar wind
    plasma with gyrotropic species pressure tensors
    may be found by neglecting the electron inertia ($m_e \equiv 0$)
    in the equations given by \citet{BarakatSchunk_82}.
Following the same procedure as given in 
    the appendix A.1 in \citet{LiLi06}, one may find
\begin{eqnarray}
&& \nabla\cdot(n_k \vec{v}_k) = 0,  \label{eq_gen_nk}  \\
&& \vec{v}_k\cdot\nabla\vec{v}_k
     +\frac{\nabla\cdot\tens{p}_k}{n_k m_k} +\frac{Z_k\nabla\cdot\tens{p}_e}{n_e m_k} 
     +\frac{GM_\odot}{r^2}\hat{r} \nonumber \\
&-& \frac{1}{n_k m_k}\left[\frac{\delta\vec{M}_k}{\delta t}
     +\frac{Z_k n_k}{n_e}\frac{\delta\vec{M}_e}{\delta t} \right] \nonumber \\
&-& \frac{Z_k}{4\pi n_e m_k}\left(\nabla\times\vec{B}\right)\times\vec{B} \nonumber \\
&+& \frac{Z_k e}{m_k c} \frac{n_j Z_j}{n_e} \left(\vec{v}_j-\vec{v}_k\right)\times\vec{B}   
    =0 ,   \label{eq_gen_vec_vk}  \\
&& \vec{v}_s\cdot\nabla p_s^{\parallel}
     +p_s^{\parallel}(\nabla\cdot\vec{v}_s + 2 \nabla_\parallel \cdot \vec{v}_s)  \nonumber \\
&+&  \nabla\cdot\vec{q}_s^{\parallel} - \tens{Q}_s\vdots\nabla(\hat{b}\hat{b})
     - \frac{\delta E_s^{\parallel}}{\delta t} = H_s^{\parallel}, \label{eq_gen_pspara}\\
&& \vec{v}_s\cdot\nabla p_s^{\perp}
     +p_s^{\perp}(\nabla\cdot\vec{v}_s + \nabla_\perp \cdot \vec{v}_s)  \nonumber \\
&+&  \nabla\cdot\vec{q}_s^{\perp} +\frac{1}{2} \tens{Q}_s\vdots\nabla(\hat{b}\hat{b})
     - \frac{\delta E_s^{\perp}}{\delta t} = H_s^{\perp}, \label{eq_gen_psperp}\\
&&\nabla\times\left(\vec{v}_e \times \vec{B}\right) = 0, \label{eq_gen_vec_magind}
\end{eqnarray}
     where the subscript $s$ refers to all species ($s=e,p,\alpha$), 
     while $k$ stands for ion species only ($k = p, \alpha$).
The gravitational constant is denoted by $G$,
     $M_\odot$ is the mass of the Sun,
     and $c$ is the speed of light.
The momentum and energy exchange rates due to the Coulomb
     collisions of species $s$ with the remaining ones
     are denoted by $\delta \vec{M}_s/\delta t$ and $\delta E_s^{\parallel, \perp}/\delta t$,
     respectively.
The third-rank tensor $\tens{Q}_s$, together with 
     the heat flux vectors $\vec{q}_s^{\parallel, \perp}$ associated with parallel and perpendicular degrees of freedom,
     arises from the deviation of species VDFs from an exact bi-Maxwellian
     \citep{BarakatSchunk_82}.
Moreover, $H_s^{\parallel, \perp}$ stands for the heating rates applied to
     species $s$ in the parallel and perpendicular directions from some non-thermal processes.
They may be determined by assuming that the heating derives from the dissipation of 
     Alfv\'en-ion cyclotron waves~\citep[e.g.,][]{HI02},
     or more simply in some ad hoc fashion such as employed in \citet{LeerAxford_72}.
The operators
     $\nabla_\parallel$ and $\nabla_\perp$ are defined by
     $\nabla_\parallel = \hat{b}\hat{b}\cdot\nabla$ and $\nabla_\perp = \nabla-\nabla_\parallel$, respectively.

In Eq.(\ref{eq_gen_vec_vk}), the subscript $j$ stands for the ion species other than $k$,
     namely, $j=p$ for $k=\alpha$ and vice versa.
As can be seen, in addition to
     the term $(\nabla\times \vec{B})\times\vec{B}$, the Lorentz force possesses 
     a new term in the form of the cross product of the ion velocity difference and
     magnetic field.
Physically, this new term represents the mutual gyration of one ion species about the other,
     the axis of gyration being in the direction of the instantaneous magnetic field.
Furthermore, Equation~(\ref{eq_gen_vec_magind}) is the time-independent version of the magnetic induction law, which
     states that the magnetic field is frozen in the electron fluid.
It may be readily shown that the effects of the electron pressure gradient, the Hall term, and the momentum exchange rates
     as contained in the generalized Ohm's law can be safely neglected given the large spatial scale in question
     (A formal evaluation of the different terms can be found in section 2.1 of \citet{Li_etal_06}).

To proceed, we choose a flux tube coordinate system, in which the base vectors are 
     $\{\hat{e}_l, \hat{e}_N, \hat{e}_\phi\}$, 
     where
\begin{eqnarray*}
\hat{e}_l = \vec{B}_P/|\vec{B}_P|, \hspace{0.5cm} \hat{e}_N = \hat{e}_\phi\times\hat{e}_l, 
\label{coor}
\end{eqnarray*}
     with the subscript $P$ denoting the poloidal component.
Moreover, the independent variable $l$ is the arclength along the
     poloidal magnetic field line measured from its footpoint at the Sun.
This choice permits the decomposition of the magnetic field
     and species velocities as follows, 
\begin{eqnarray}
\vec{B} = B_l \hat{e}_l + B_\phi \hat{e}_\phi,\hspace{0.5cm}
\vec{v}_s = v_{sl}\hat{e}_l + v_{sN}\hat{e}_N + v_{s\phi}\hat{e}_\phi ,
\label{vec_BV_comps}
\end{eqnarray}
     where $s=e, p, \alpha$. 
From the assumption of azimuthal symmetry, and the assumption that 
     the solar wind is time-independent,
     one can see from the poloidal component of equation~(\ref{eq_gen_vec_magind})
     that $\vec{v}_{eP}$ should be strictly in the direction of $\vec{B}_P$.
In other words, $v_{e N}=0$ to a good approximation.
Now let us consider the $\phi$ component of the momentum equation~(\ref{eq_gen_vec_vk}).
Since the  frequencies associated with the spatial dependence
     are well below the ion gyro-frequency
     $\Omega_k = (Z_k e B_l)/(m_k c)$ ($k=p, \alpha$), 
     from an order-of-magnitude estimate one can see that
     $|v_{j N}-v_{k N}| \ll |v_{k \phi}|$.
Combined with the fact that $v_{e N}=0$, this leads to that
     both $v_{p N}$ and $v_{\alpha N}$ should be very small and 
     can be safely neglected unless they appear alongside the
     ion gyro-frequency.
With this in mind, one can find from the $N$ component of equation~(\ref{eq_gen_vec_vk}) that
\begin{eqnarray}
v_{\alpha \phi}-v_{p \phi} = \frac{B_\phi}{B_l}\left(v_{\alpha l}-v_{p l}\right). 
\label{ion_vdiff}
\end{eqnarray}
That is, the ion velocity difference is strictly aligned with the 
     magnetic field.
This alignment condition further couples one ion species to the other.

The fact that $v_{s N}$ ($s=e, p, \alpha$) is negligible means that
     the system of vector equations may be decomposed into 
     a force balance condition across the poloidal magnetic field
     and a set of transport equations along it.
In the present paper, however, we simply replace the force balance condition by
      prescribing an analytical meridional magnetic field configuration.
Moreover, we examine in detail only the azimuthal dynamics, leaving
     a brief discussion on the poloidal one in the appendix.

\subsection{Azimuthal dynamics}
\label{sec_mathform_azi}

The $\phi$ component of the magnetic induction law (\ref{eq_gen_vec_magind}) gives
\begin{eqnarray}
 \nabla\cdot \left[\frac{1}{R}\left(B_\phi \vec{v}_{e P} - v_{e \phi} \vec{B}_{P}\right)\right] =0.
\label{eq_red_magind}
\end{eqnarray}
Now that $\vec{v}_{eP} = v_{e l} \hat{e}_l$, one may readily integrate Eq.(\ref{eq_red_magind}) along
    a magnetic line of force
    to yield
\begin{eqnarray}
 v_{e \phi} = A_\Omega R + \frac{B_\phi}{B_l} v_{e l}.
\end{eqnarray}
Here $R=r\sin\theta$ is a geometrical factor to be evaluated along a given line of force (see Fig.\ref{fig_mf}),
    and $A_\Omega$ is a constant of integration and should be identified as
    the angular rotation rate of the footpoint of the magnetic flux tube.
Taking into account the alignment condition~(\ref{ion_vdiff}), one may find that
\begin{eqnarray}
 v_{s \phi} = A_\Omega R + \frac{B_\phi}{B_l} v_{s l}
\label{eq_align_all}
\end{eqnarray}
    where $s=e, p, \alpha$.
Therefore in a frame of reference that corotates with the Sun, the velocities of all species 
    are aligned with the magnetic field.

Another equation that enters into the azimuthal dynamics is the $\phi$ component
    of the total momentum.
In the present case, it reads
\begin{eqnarray}
&& \frac{1}{R}\left\{\sum_k \rho_k v_{k l} \left(R v_{k \phi}\right)' \right. \nonumber \\
&&  - \left.\frac{B_l}{4 \pi}\left[\left(1-\frac{4\pi P^\Delta}{B^2}\right)R B_\phi\right]'\right\} =0 ,
\end{eqnarray}
    where
\begin{eqnarray}
P^{\Delta} = P^{\parallel} - P^{\perp}, \hspace{0.2cm} 
   P^{\parallel, \perp} = \sum_s p_s^{\parallel, \perp} , 
\end{eqnarray}
    and the prime $'=\hat{e}_l\cdot\nabla$ is the directional derivative along
    the poloidal magnetic field.

For a time-independent flow $\rho_k v_{k l}/B_l = \mbox{const}$.
It then follows that 
\begin{eqnarray}
R\left[ v_{p \phi} + \eta v_{\alpha \phi} - 
 \frac{B_l B_\phi}{4\pi \rho_p v_{pl}}\left(1-\frac{4\pi P^\Delta}{B^2}\right)\right] = A_L,
\label{eq_ang_cons}
\end{eqnarray}
     where the constant $\eta = (\rho_\alpha v_{\alpha l})/(\rho_p v_{p l})$ is the ion
     mass flux ratio, and $A_L$ is a constant of integration.
Physically, $A_L$ is related to the angular momentum loss rate
     per steradian $L$ by 
\begin{eqnarray}
 L = \dot{M}_p A_L, 
\end{eqnarray}
    where
\begin{eqnarray}
  \dot{M}_k =  \rho_k v_{k l}\frac{B_{l E}}{B_l} r_E^2,\hspace{0.5cm} (k=p, \alpha)
\label{eq_def_dotMp}
\end{eqnarray}
     is the ion mass loss rate per steradian scaled to the Earth orbit $r_E=1$~AU,
     with $B_{l E}$ denoting the strength of the poloidal magnetic field at $r_E$.
It follows that the angular momentum loss rate of the Sun due to the solar wind $\dot{L}=4\pi L$ 
     if $L$ is independent of colatitude.
Equation~(\ref{eq_ang_cons}) shows that $L$ consists of the contributions due to
     individual ion angular momenta $L_k$, the magnetic stresses $L_M$ and
     the total pressure anisotropy $L_{ani}$, where
\begin{eqnarray}
&& \left[L_k, L_M, L_{ani}\right] \nonumber \\
&=& R\left(\frac{B_{lE}}{B_l}r_E^2\right) 
 \left[\rho_k v_{k l}v_{k\phi}, \frac{-B_l B_\phi}{4\pi}, \frac{B_l B_\phi P^\Delta}{B^2}\right]
\label{eq_def_Ls}
\end{eqnarray}
with  $k=p, \alpha$.

Substituting Eq.(\ref{eq_align_all}) into (\ref{eq_ang_cons}), one may find
\begin{eqnarray}
 &&  \tan\Phi \left[M_T^2 - \left(1-\beta^\Delta\cos^2\Phi\right)\right] \nonumber \\
 &=&  \epsilon \left[\frac{A_L}{(1+\eta)A_\Omega R^2}-1\right] ,
\label{eq_tanp_prelim}
\end{eqnarray}
      where $\tan\Phi = B_\phi/B_l$ defines the magnetic azimuthal angle $\Phi$, and
\begin{eqnarray}
&& M_T^2 = M_p^2 + M_\alpha^2, 
 M_k^2 = \frac{4\pi \rho_k v_{k l}^2}{B_l^2}\hspace{0.2cm}(k=p, \alpha), \nonumber \\
&& \beta^\Delta = \frac{4\pi P^\Delta}{B_l^2}, \hspace{0.2cm}
   \epsilon = (1+\eta) M_p^2 \frac{A_\Omega R}{v_{p l}}.
\label{eq_def_MT2_betaDelta}
\end{eqnarray}
By definition, $M_T$ is the combined poloidal Alfv\'enic Mach number, which involves 
     both ion species.
For a typical solar wind, between 1~$R_\odot$ and 1~AU there exists a point where $M_T=1$,
     which is to be called the Alfv\'en point and denoted by $r_a$.

As discussed in detail by \citet{LiLi06}, when species temperature anisotropy is absent
     ($P^\Delta=0$ and therefore $\beta^\Delta=0$), for Eq.(\ref{eq_tanp_prelim}) to possess
     a solution that passes smoothly through $r_a$ the two constants 
     $A_L$ and $A_\Omega$ have to be related by
\begin{eqnarray}
 A_L = (1+\eta) A_\Omega R_a^2 ,
\label{eq_L_isotropic}
\end{eqnarray}
     where the subscript $a$ denotes quantities evaluated at the Alfv\'en point.
When $\beta^\Delta$ is not zero, a direct relation between $A_L$ and $A_\Omega$ is not as obvious since
     now Eq.(\ref{eq_tanp_prelim}) becomes cubic in $\tan\Phi$.
Nevertheless, one may write $A_L$ as $A_L = \lambda A_{L 0}$, where $A_{L 0}$ is determined through Eq.(\ref{eq_L_isotropic})
     and therefore $\lambda$ stands for the correction due to a finite $\beta^\Delta$.
It then follows that 
\begin{eqnarray}
 c_3 \tan^3\Phi +  c_2 \tan^2\Phi +  c_1 \tan\Phi + c_2 = 0,
\label{eq_tanp_standard}
\end{eqnarray}
where 
\begin{eqnarray}
&&  c_3 = M_T^2-1, c_2 = \epsilon\left[1-\lambda\left(\frac{R_a}{R}\right)^2\right],\nonumber \\
&&  c_1 = M_T^2-1+\beta^\Delta.
\end{eqnarray}
Given the meridional flow profiles along a prescribed magnetic field line,
      Eq.(\ref{eq_tanp_standard}) possesses only one real root at locations
      far away from $r_a$.
However, in the vicinity of $r_a$, there exists in general three real roots and they diverge
      near $r_a$.
The requirement that there exists a unique solution that is smooth from 1~$R_\odot$ out to 1~AU 
      determines $\lambda$~\citep{WD_70,Weber_70}.


\section{Meridional magnetic field and flow profiles}
\label{sec_presflow}
In principle, one needs to solve Eqs.(\ref{eq_reduced_nk}) to (\ref{eq_reduced_Tsper}) together with
     Eq.(\ref{eq_tanp_prelim}) simultaneously to gain a 
     a quantitative insight.
In the present paper, we refrain from doing so because from previous experience it
     proves difficult to yield the flow profiles that satisfactorily reproduce
     in situ measurements such as made by Helios.
Take the proton-alpha speed difference $v_{\alpha p}$ in the fast solar wind for example.
It is observationally established that $v_{\alpha p}$ closely tracks the local Alfv\'en speed
     in the heliocentric range $r> 0.3$~AU~\citep{Marsch_etal_82a}.
So far this fact still poses a theoretical challenge: adjusting the ad hoc heating parameters,
     or fine-tuning the cyclotron resonance mechanism is unable to
     produce such a behavior~\citep[see, e.g.][]{HuHabbal_99}.
We therefore adopt an alternative approach by prescribing the background meridional flow profiles
     that mimic the observations and then examining what consequences
     the species anisotropies have on the azimuthal dynamics.

\subsection{Background meridional magnetic field}
\label{sec_presflow_bkgrndMF}

For the meridional magnetic field, we adopt an analytical model given by 
     \citet{Bana_etal_98}.
In the present implementation, the model magnetic field consists of
     the dipole and current-sheet components only.
A set of parameters $M=2.2265$, $Q=0$, $K=0.9343$ and $a_1=1.5$ are
     chosen such that the last open magnetic field line is anchored
     at heliocentric colatitude $\theta=50^\circ$ on the Sun,
     while at the Earth orbit, the meridional magnetic field strength
     $B_l$ is 3$\gamma$ and independent of colatitude $\theta$, 
     consistent with Ulysses measurements~\citep{SmithBalogh_95}.

\begin{figure}
 \centering
  \includegraphics[width=0.49\textwidth]{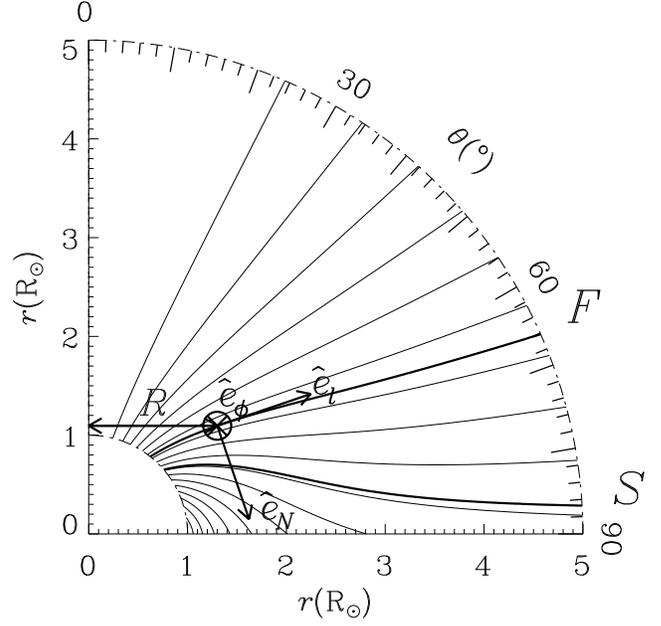}
 \vskip 0.5cm
 \caption{
Adopted meridional magnetic field configuration in the inner corona.
Here only a quadrant is shown in which the magnetic axis points upward, and the thick contours 
     labeled $F$ and $S$ delineate the lines of force along which the fast
     and slow solar wind solutions are examined, respectively.
Also shown is how to define the geometrical factor $R$,
     and the base vectors $\hat{e}_l, \hat{e}_N$ and $\hat{e}_\phi$ 
     of the flux tube coordinate system
     (see section~\ref{sec_mathform}).
  }
\label{fig_mf}
\end{figure}

The background magnetic field configuration is depicted in Fig.\ref{fig_mf},
     where the thick contours labeled $F$ and $S$
     represent the lines of force along which we examine the
     fast and slow solar wind solutions, respectively.
Tube $F$ ($S$), which intersects the Earth orbit at 70$^\circ$ (89$^\circ$) colatitude,
     originates from $\theta=38.5^\circ$ ($49.4^\circ$)
     at the Sun where the meridional magnetic field strength $B_l$ is
     $3.93$ ($3.49$)~G.

\subsection{Prescribed meridional flow profiles}
\label{sec_presflow_bkgrndFlow}

\begin{figure*}
 \centering
  \includegraphics[width=.98\textwidth]{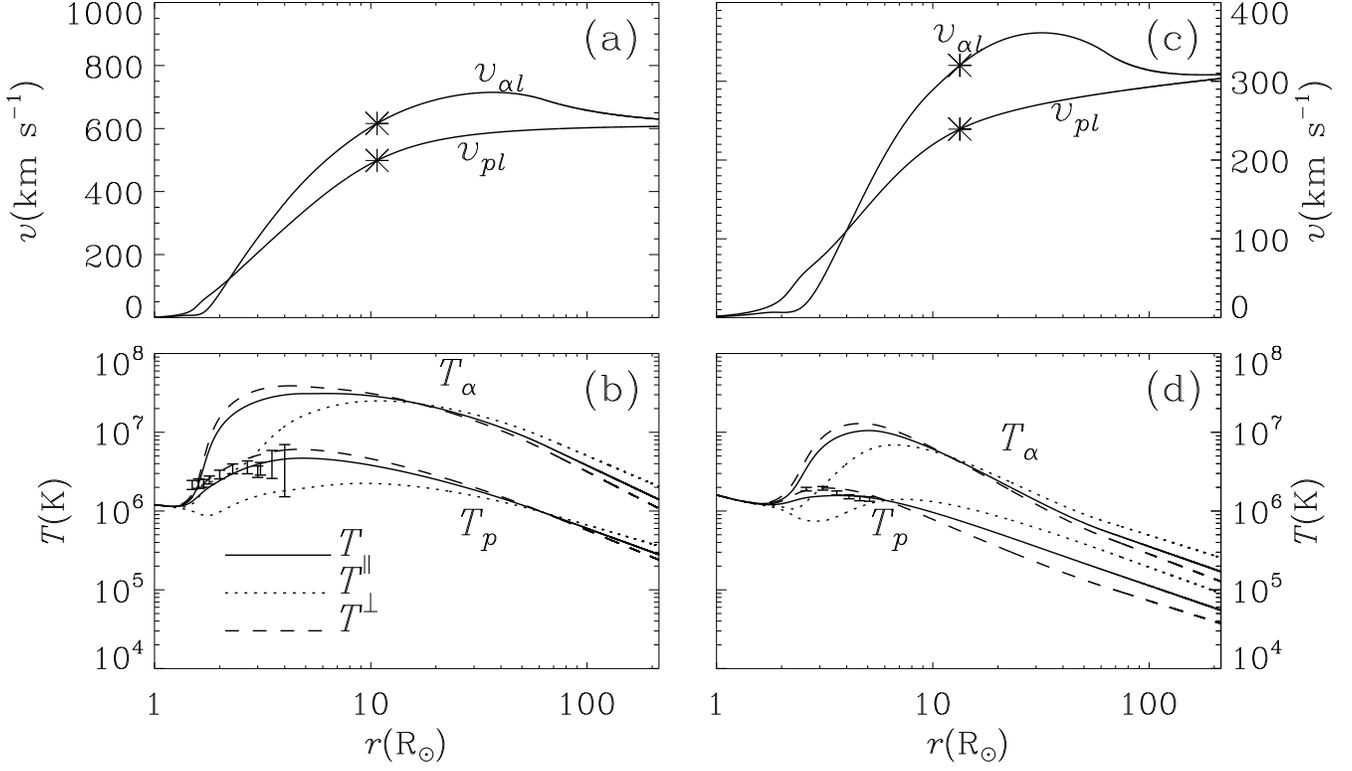}
 \vskip 0.5cm
 \caption{
Radial distribution between 1~$R_\odot$ and 1~AU of the adopted meridional flow parameters for the fast (left column) and slow (right) solar wind.
(a) and (c), the meridional flow speeds of protons ($v_{pl}$) and alpha particles ($v_{\alpha l}$).
(b) and (d), the ion temperatures $T_k^{\parallel}$ (dotted lines), $T_k^{\perp}$ (dashed lines),
     and $T_k=(T_k^{\parallel}+2T_k^{\perp})/3$ (solid lines) where $k=p, \alpha$.
The construction of $T_k^{\parallel, \perp}$ is described in section~\ref{sec_presflow_bkgrndFlow}.
The error bars in (b) and (d) represent the uncertainties of the UVCS measurements
    of the effective proton temperature as reported by \citet{Kohl_etal_98} for a coronal hole,
    and by \citet{Frazin_etal_03} for a streamer, respectively.
Note that both measurements are typical of solar minimum conditions.
Moreover, the asterisks in (a) and (c) denote the Alfv\'en point, where the meridional
    Alfv\'enic Mach number (defined by Eq.(\ref{eq_def_MT2_betaDelta})) equals unity.
  }
\label{fig_bkgrn_flow}
\end{figure*}

The background meridional flow parameters are found by adopting a three-step approach described as follows:
\begin{enumerate}
 \item 
Using some ad hoc heating parameters, we solve along flux tube $F$ ($S$) the {\it isotropic} 
     version of Eqs.(\ref{eq_reduced_nk}) to (\ref{eq_reduced_Tsper}) (see Eqs.(8) to (10) in \citet{LiLi07} for details)
     to yield the distribution between 1~$R_\odot$ and 1~AU of the ion number densities $n_k$ and meridional
     speeds $v_{k l}$ ($k=p, \alpha$), as well as the {\it isotropic}
     species temperatures $T_s$ ($s=e, p, \alpha$)
     for the fast (slow) solar wind.
Specifically, the heating rates are of the same format as in section 3.2 in \citet{LiLi08}.
     To generate the fast and slow solar wind solutions, 
     the parameters $[F_E$~(in\funits), $l_d$~(in\lengthunits), $\chi]$ are chosen to be $[1.9, 2.2, 2.2]$
     and $[1, 1.8, 3.7]$, respectively.
By simply adjusting the heating parameters it proves difficult to produce a reasonable $T_p$ profile in that 
     if $T_p$ in the inner corona is close to observations then $T_p$ at 1~AU is usually only a fraction of
     the typically measured values.
Moreover, the derived speed difference $v_{\alpha p, l}=v_{\alpha l}-v_{p l}$ varies little between 0.3 and 1~AU,
     in contrast to the Helios measurements.
Therefore some additional steps are employed to make the flow profiles more realistic.
Specifically, all the parameters except $n_p$ and $v_{pl}$ are required to undergo a smooth transition from
     the profiles for the region $r\lesssim 0.3$~AU derived so far
     to those specified in next step for the outer region.

\item 
The desired profiles, given in Table~\ref{tab_outer}, for $v_{\alpha p, l}$, $T_p$ and $T_\alpha$ for the region $r\gtrsim 0.3$~AU
          (denoted by subscript $o$)
          are based on the in situ measurements to be detailed shortly. 
Once $v_{\alpha p, l}$ is known, the meridional alpha speed $v_{\alpha l}$ is given by $v_{p l}+v_{\alpha p, l}$,
      and the alpha density $n_\alpha$ by $(n_\alpha)_I (v_{\alpha l})_I/v_{\alpha l}$,
      where the subscript $I$ denotes the values obtained in the first step.
The distributions of $n_k$, $v_{kl}$ ($k=p, \alpha$) and $T_s$ ($s=e, p, \alpha$) thus constructed are for the
      isotropic model.

\item 
Now the ion temperatures $T_k^{\parallel, \perp}$ can be constructed by prescribing the temperature anisotropy
      $\Gamma_k = T_k^{\parallel}/T_k^{\perp}$ ($k=p, \alpha$).
Note that the electron temperature is assumed to be isotropic.
For the region within several solar radii, $\Gamma_k$ is required to decrease with $r$ from $1$ at 1~$R_\odot$, 
      where the Coulomb self-collisions are still frequent enough to suppress a temperature anisotropy, to some value less than unity.
This inner profile is not directly constrained by observations but constructed by noting that 
      the processes operational in the inner corona tend to heat the ions preferentially in the perpendicular direction
      \citep[e.g.,][]{HI02}.
On the other hand, for $r \gtrsim 0.3$~AU, $\Gamma_k$ follows a power law dependence on $r$ with the exponent
      determined by the Helios measurements (see Table~\ref{tab_outer}).
Specifying the temperature anisotropies of protons and alpha particles at 1~AU, $\Gamma_{p E}$ and $\Gamma_{\alpha E}$, 
      determines $\Gamma_k$ in the outer region. 
The inner and outer profiles are then connected smoothly to yield the desired $\Gamma_k$.
The temperatures $T_k^{\parallel, \perp}$ follow from the relations
     $T_k^{\perp} = 3T_k/(2+\Gamma_k)$ and $T_k^{\parallel} = \Gamma_k T_k^{\perp}$.
\end{enumerate}

A detailed description of Table~\ref{tab_outer} is necessary.
Note that throughout this table $x=r/r_E$ where $r_E=1$~AU.
Let us first focus on the adopted values for the fast solar wind.
For the {\it isotropic} proton and alpha temperatures at 1~AU, we adopted
    the typical values of $T_p=2.8\times 10^5$~K and $T_\alpha=5 T_p$~\citep[see e.g.,][]{Schwenn_90,McComas_etal_00}
    (hereafter Sch90 and Mc00).
Furthermore, Figures~18 and 19 in~\citet{Marsch_etal_82b} (hereafter M82b) indicate
    that $T_p^\parallel \propto x^{-0.75}$, and $T_p^{\perp} \propto x^{-1.08}$. 
A power law dependence for $T_p$ of $T_p\propto x^{-1}$ is therefore consistent with such a behavior, and also consistent with
    the Ulysses measurements (see Table 2 in~Mc00).
Furthermore, Figure~5 in~\citet{Marsch_etal_82a} (hereafter M82a) indicates
     that $T_\alpha^\parallel \propto x^{-1.15}$,
     and $T_\alpha^{\perp} \propto x^{-1.38}$. 
A profile of $T_\alpha \propto x^{-1.3}$ is consistent with this behavior,
    but differs substantially from that measured by Ulysses, 
    which yields that $T_\alpha\propto x^{-0.8}$ (see Table 2 in Mc00).
Moving on to the slow solar wind, we note that 
    values of $T_p=5.5\times 10^4$~K and $T_\alpha=1.7\times 10^5$~K are 
    typically found at 1~AU (see e.g., Sch90).
In addition, Figures~18 and 19 in M82b indicate that $T_p^\parallel \propto x^{-1.03}$,
    and $T_p^{\perp} \propto x^{-0.9}$. 
Therefore we adopted a $T_p$ profile of $T_p\propto x^{-0.94}$.
On the other hand, we adopted a profile for $T_\alpha$ in the form $T_\alpha \propto x^{-0.96}$, which is consistent with
     the measured alpha temperature anisotropies which indicate that 
     $T_\alpha^\parallel \propto x^{-0.83}$,
     and $T_\alpha^{\perp} \propto x^{-1.02}$ (see Fig.5 in M82a).

In this study $\Gamma_{p E}$ and $\Gamma_{\alpha E}$ will serve as free parameters. 
The Helios measurements indicate that $\Gamma_{p E} \approx 1.2\pm 0.3$ and $\Gamma_{\alpha E} \approx 1.3\pm 0.6$
    for the fast solar wind with $v_{pl} \gtrsim 600$\velunits,
    while $\Gamma_{p E} \approx 1.7\pm 0.7$ and $\Gamma_{\alpha E} \approx 1.4\pm 0.6$ for the slow solar
    wind with $v_{p l} \lesssim 400$\velunits~\citep{Marsch_etal_82a, Marsch_etal_82b}.
Theoretically, one may expect that the $[\Gamma_{p E}, \Gamma_{\alpha E}]$ pair may not
    occupy the whole rectangle bounded by the given values in the $\Gamma_{p E}$-$\Gamma_{\alpha E}$ space, since
    too strong an anisotropy can drive the system unstable with respect to a number of instabilities
    when the plasma $\beta$ is comparable to unity. 
Given that the lower limit of $\Gamma_{p E}$ or $\Gamma_{\alpha E}$ is only slightly lower than $1$, the ion-cyclotron instability
    can be shown to be unlikely to occur \citep[see, e.g., Eq.(3) in][]{Gary_etal_94}.
However, the firehose instability may be relevant since it happens when $P^\parallel$ is sufficiently larger than $P^\perp$ and
    $\beta^\parallel =8\pi P^\parallel/B^2 \gtrsim 1$.
Note that the alpha particles with a non-negligible abundance drifting relative to protons may complicate the situation
    considerably given that in addition to the firehose, electromagnetic ion/ion instabilities may also be relevant and the occurrence
    of such instabilities is not restricted to the cases where the parallel $\beta$ is large~\citep{HellingerTravnicek_06}.
Nevertheless, we only compare the modeled $[\Gamma_p, \Gamma_\alpha]$ with the non-resonant firehose
    criterion such as found via the dispersion relation of Alfv\'en waves \citep[see Eq.(23) in][]{I84}.
Specializing to an electron-proton-alpha plasma, the dispersion relation dictates that instability occurs when 
    $1 - P^\perp/P^\parallel > 2 (1-x_p x_\alpha)/\beta^\parallel$ where $x_k = (\rho_k/\rho) (v_{\alpha p}/v_A)$ ($k=p, \alpha$)
    with $v_A=B/\sqrt{4\pi\rho}$ being the Alfv\'en speed determined by the bulk mass density $\rho = \rho_p+\rho_\alpha$.
Using this criterion it is found that the modeled flow profiles are all stable with the only exception being for the segment
    $r \gtrsim 195$~$R_\odot$ in the fast wind with the largest values of $\Gamma_{pE}$ and $\Gamma_{\alpha E}$.

Figure~\ref{fig_bkgrn_flow} gives the radial distributions between 1~$R_\odot$ and 1~AU
    of the flow parameters for the fast and slow solar wind in the left and right panels, respectively.
Figures~\ref{fig_bkgrn_flow}a and \ref{fig_bkgrn_flow}c depict
    the meridional ion speeds $v_{pl}$
    and $v_{\alpha l}$, while 
    the ion temperatures $T_k^{\parallel}$ (the dotted curves), $T_k^\perp$ (dashed) and $T_k$ (solid) 
    are given in Figs.\ref{fig_bkgrn_flow}b and \ref{fig_bkgrn_flow}d ($k=p, \alpha$).
The values for the temperature anisotropy adopted for the construction are
    $\Gamma_{p, E}=1.5$ and $\Gamma_{\alpha, E}=1.9$ for the fast wind, 
    and $\Gamma_{p, E}=2.42$ and $\Gamma_{\alpha, E}=2$ for the slow wind.
In Fig.\ref{fig_bkgrn_flow}b, the error bars represent the uncertainties
    of the UVCS measurements for the proton effective temperature, made for a polar coronal hole as reported by \citet{Kohl_etal_98}.
Similar measurements by \citet{Frazin_etal_03} along the edges of an equatorial streamer
    are given in Fig.\ref{fig_bkgrn_flow}d.
Moreover, the asterisks in Figs.\ref{fig_bkgrn_flow}a and \ref{fig_bkgrn_flow}c mark the location of the Alfv\'en point as defined by
    Eq.(\ref{eq_def_MT2_betaDelta}).

For the fast (slow) solar wind it is found that at 1~AU the meridional proton speed $v_{pl}$ is $607$ ($304$)\velunits,
     the proton flux $n_p v_{pl}$ is $2.8$ ($3.84$) in units of $10^{8}$\noflxunits,
     the alpha abundance $n_{\alpha}/n_p$ is 4.56\% (3.6\%),
     and the meridional component of the proton-alpha velocity difference $v_{\alpha p, l}$ is 23 (5)\velunits.
These values are consistent with in situ measurements such as made by Ulysses~\citep{McComas_etal_00}.
Moreover, the fast (slow) solar wind reaches the Alfv\'en point at 10.7 (13.3)~$R_\odot$, beyond which
     $v_{pl}$ increases only slightly with increasing $r$.
On the other hand, for $r\gtrsim 0.3$~AU the meridional alpha speed $v_{\alpha l}$ decreases
     rather than increases with $r$ as a consequence of the prescribed $v_{\alpha p, l}$ profile.
If examining the ratio of $v_{\alpha p, l}$ to the meridional Alfv\'en speed $v_{Al} = B_l/\sqrt{4\pi\rho}$,
     one may find that for the fast solar wind this ratio decreases only slightly from $0.98$ at 0.3~AU to $0.82$ at 1~AU,
     while for the slow wind it shows a substantial variation from $0.88$ at 0.3~AU to $0.29$ at 1~AU.
The modeled $v_{\alpha p, l}/v_{Al}$ can be seen to agree with the Helios measurements as given by Fig.11 of \citet{Marsch_etal_82a}.
Note that a value of $v_{\alpha p, l}=49$\velunits\ at 0.3~AU is 
       not unrealistic for slow solar winds, even larger values have been found
       by Helios 2 when approaching perihelion~\citep{Marsch_etal_81}.
Moving on to the temperature profiles, one may see that the $T_p^{\perp}$ profiles inside 5~$R_\odot$
       are in reasonable agreement with the UVCS line-width measurements
       for both the fast and slow solar wind.
\begin{table}
\begin{minipage}[t]{\columnwidth}
\caption{Profiles for some solar wind parameters in the region $r> 0.3$~AU.}
\label{tab_outer}
\centering
\renewcommand{\footnoterule}{}  
\begin{tabular}{lcc}
\hline\hline
			& fast wind & slow wind \\
\noalign{\smallskip}
\hline
\noalign{\smallskip}
$(v_{\alpha p, l})_o$	
& $23 x^{-1.15}$\velunits~\footnote{please see section 3.2 for details} 
& $5 x^{-1.9}$\velunits\\
\noalign{\smallskip}
\hline
\noalign{\smallskip}
$(T_{p})_o$		
&  $2.8\times 10^5 x^{-1}$~K
&  $5.5\times 10^4 x^{-0.94}$~K \\
$(T_{\alpha})_o$ 
&  $1.4\times 10^6 x^{-1.3}$~K
&  $1.7\times 10^5 x^{-0.96}$~K \\
\noalign{\smallskip}
\hline
\noalign{\smallskip}
$(\Gamma_{p})_o$		
&   $\Gamma_{p E} x^{0.33}$
&   $\Gamma_{p E} x^{-0.13}$\\

$(\Gamma_{\alpha})_o$   	
&   $\Gamma_{\alpha E} x^{0.23}$	
&   $\Gamma_{\alpha E} x^{0.19}$ \\ 	

\noalign{\smallskip} \hline
\end{tabular}
\end{minipage}
\end{table}

\section{Numerical results}
\label{sec_numres}
Having described the meridional magnetic field and flow profiles, we may now address
     the following questions: to what extent is the total angular momentum loss of the Sun affected
     by the ion temperature anisotropies? and how is the angular momentum budget
     distributed among particle momenta, the magnetic torque, and the torque due to ion temperature anisotropies?
To this end, let us first examine the fast and then the slow solar wind solutions.
In the computations, we take $A_\Omega=2.865\times 10^{-6}$~rad~s$^{-1}$, which corresponds to
    a sidereal rotation period of $25.38$~days.

\subsection{Fast solar wind}
\label{sec_numres_fast}

\begin{figure}
 \centering
  \includegraphics[width=.49\textwidth]{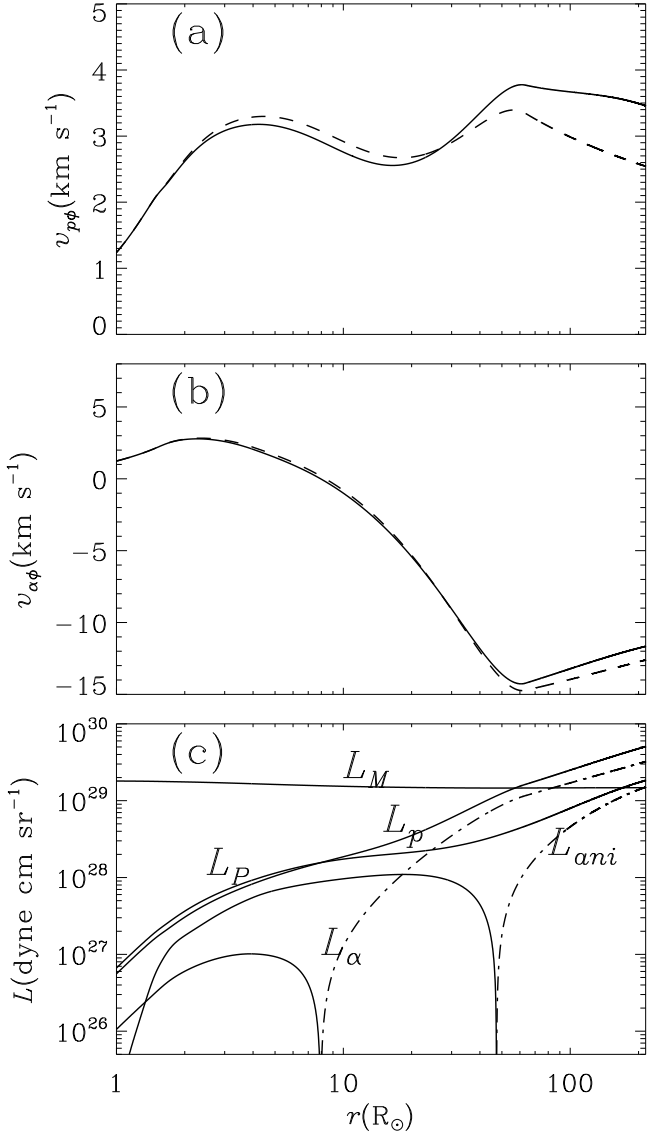}
 \vskip 0.5cm
 \caption{
Radial distributions of (a) the proton azimuthal speed $v_{p\phi}$, (b) the alpha one $v_{\alpha\phi}$, and
     (c) various contributions to the angular momentum budget in an $e-p-\alpha$ solar wind with ion temperature anisotropies.
In (a) and (b), the profiles derived for a solar wind with identical flow parameters where ion temperature anisotropies are neglected
    are given by dashed lines for comparison.
Panel (c) depicts the individual ion angular momentum fluxes $L_p$ and $L_\alpha$, their sum $L_P$,
        and the fluxes associated with the magnetic stresses $L_M$, 
	and with the temperature anisotropies $L_{ani}$ (see Eq.(\ref{eq_def_Ls})). 
The dash-dotted lines represent negative values.
  }
\label{fig_angmom_fast_case}
\end{figure}

Figure~\ref{fig_angmom_fast_case} presents the radial profiles of (a) the proton azimuthal speed $v_{p\phi}$,
	(b) the alpha one $v_{\alpha\phi}$,
        and (c) the ion angular momentum fluxes $L_k$ ($k=p, \alpha$), 
	their sum $L_P$,
        the flux due to the magnetic torque $L_M$,
        and that due to temperature anisotropies $L_{ani}$
	(see Eq.(\ref{eq_def_Ls})).
Note that the dash-dotted curves in Fig.\ref{fig_angmom_fast_case}c plot negative values.
In Figs.\ref{fig_angmom_fast_case}a and \ref{fig_angmom_fast_case}b, the ion azimuthal speeds
	for the isotropic model with identical meridional flow parameters
	are given by the dashed lines for comparison.
The fast wind profile corresponds to $\Gamma_{p E}=1.5$ and $\Gamma_{\alpha E}= 1.9$.

For the chosen $\Gamma_{pE}$ and $\Gamma_{\alpha E}$, it is found that $\lambda=1.058$.
Consequently, the total angular momentum loss rate per steradian $L$ is $1.8$
         (here and hereafter in units of $10^{29}$\specangunits) 
         in the anisotropic case, and is only modestly enhanced compared with the isotropic case, for which
         $L=1.71$.
Furthermore, Figs.\ref{fig_angmom_fast_case}a and \ref{fig_angmom_fast_case}b indicate
	 that the radial dependence of the ion azimuthal speed $v_{p\phi}$ or $v_{\alpha\phi}$
	 in the anisotropic model is similar to that in the isotropic one.
For instance, both models yield that with increasing distance the alpha particles
         develop an azimuthal speed in the direction of counterrotation with the Sun: 
         $v_{\alpha \phi}$ becomes negative beyond $7.95$ ($8.35$)~$R_\odot$ in the
         anisotropic (isotropic) model.
The difference between the isotropic and anisotropic cases
         becomes more prominent at large distances where $\beta^\Delta$
         becomes increasingly significant, as would be expected from Eq.(\ref{eq_tanp_prelim}).
Take the values of $v_{p\phi}$ and $v_{\alpha\phi}$ at 1~AU.
The isotropic (anisotropic) model yields that $v_{p\phi}=2.54$ ($3.46$)\velunits\ 
        and that $v_{\alpha\phi}=-12.6$ ($-11.7$)\velunits\ at 1~AU.
Note that the changes introduced to the ion azimuthal speeds by pressure anisotropies 
        ($\sim 0.9$\velunits\ for both protons and alpha particles)
        play an important role in the distribution of the angular momentum budget $L$
	among different contributions, as shown by Fig.\ref{fig_angmom_fast_case}c.
The proton contribution $L_p$ exceeds $L_M$ for $r\gtrsim 57$~$R_\odot$
        and $L_p$ attains $5.07$ at 1~AU, significantly larger than the magnetic part $L_M=1.48$ at the same location.
In fact, the overall particle contribution $L_P$, which increases with distance,
        overtakes the magnetic contribution $L_M$ from $170.5$~$R_\odot$ onwards,
        despite the fact that the alpha contribution tends to
	offset the proton one.
The dominance of $L_P$ over $L_M$ happens in conjunction with the increasing importance of
        $L_{ani}$, the flux due to total pressure anisotropy which is in the direction of counterrotation with the Sun.
In contrast, without pressure anisotropies, at 1~AU it turns out that even though a value of $3.73$ is found for $L_p$, 
        it is almost cancelled by an $L_\alpha$ of $-3.49$.
The resulting $L_P$ is thus $0.23$, substantially smaller than $L_M$, which is nearly identical to the value found
        in the anisotropic model.
This contrast between anisotropic and isotropic cases is understandable since it follows from Eq.(\ref{eq_ang_cons}) that,
        given that the constant
        $A_L$ does not vary much from the isotropic to anisotropic model,
        the change of $L_P$ should be largely offset by that of $L_{ani}$.

\begin{figure}
 \centering
  \includegraphics[width=.49\textwidth]{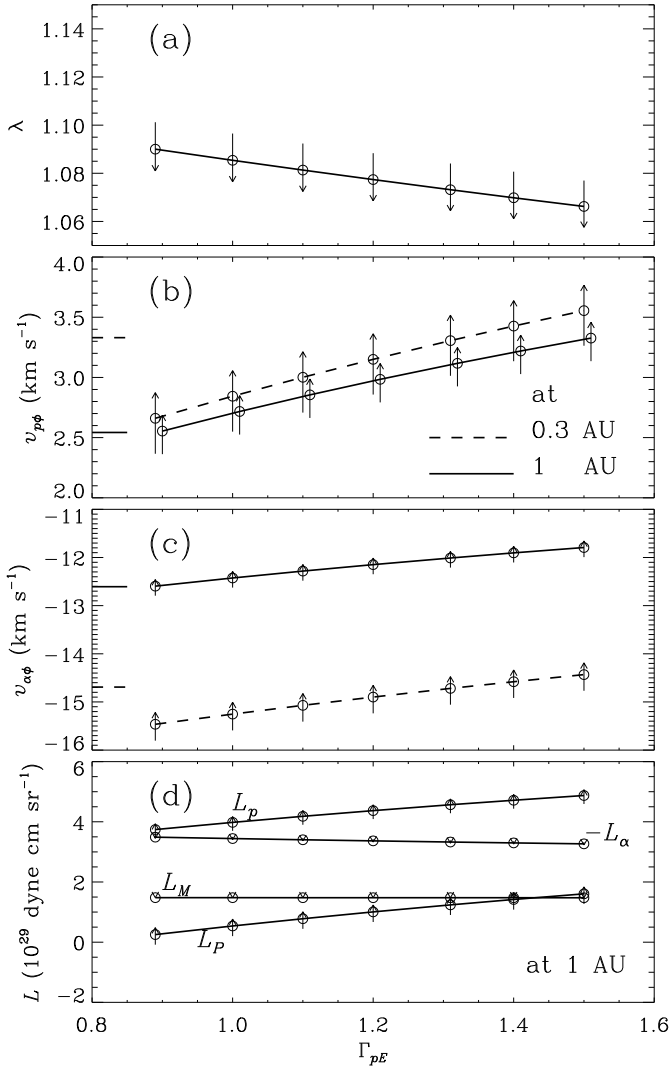}
 \vskip 0.5cm
 \caption{
Values of several parameters as a function of $\Gamma_{pE}$, the proton temperature anisotropy at 1~AU.
(a) The factor $\lambda$, the deviation of which from unity represents the correction to the total angular momentum
        loss due to the introduction of ion pressure anisotropies;
(b) and (c) the proton and alpha azimuthal speeds $v_{p\phi}$ and $v_{\alpha\phi}$ at two different heliocentric distances
        given by different line styles as indicated in (b);
(d) various components in the angular momentum flux at 1~AU including individual ion contribution $L_p$ and $L_\alpha$,
        the overall particle contribution $L_P=L_p+L_\alpha$,
        as well as the contribution from magnetic stresses $L_M$.
Note that $-L_\alpha$ instead of $L_\alpha$ is given in (d). 
The short horizontal bars in panels (b) and (c) represent the azimuthal ion speeds derived in the isotropic model for
          comparison.
Furthermore, in panel (b) the curve corresponding to 1~AU is slightly shifted relative to that for 0.3~AU to avoid the
        two overlapping each other.
The open circles correspond to the cases where $\Gamma_{\alpha E}$ is fixed at $1.3$, where
      $\Gamma_{\alpha E}$ is the alpha temperature anisotropy at 1~AU.
At a given $\Gamma_{pE}$ each parameter varies monotonically from the value with $\Gamma_{\alpha E}=0.7$, represented by the end of 
      the arrow, to the value with $\Gamma_{\alpha E}=1.9$ given by the arrow head.
The ranges in which $\Gamma_{pE}$ and $\Gamma_{\alpha E}$ vary are determined from the Helios measurements
	(see text for details).
  }
\label{fig_angmom_fast_par}
\end{figure}

Figure~\ref{fig_angmom_fast_par} expands the obtained results by displaying the dependence on $\Gamma_{pE}$
        and $\Gamma_{\alpha E}$ of (a) the factor $\lambda$, 
	(b) the proton azimuthal speed $v_{p\phi}$ and (c) the alpha one $v_{\alpha\phi}$ at two different distances
        plotted by the different linestyles indicated in (b),
        as well as (d) the constituents comprising the angular momentum flux at 1~AU.
In addition to the individual ion contributions $L_p$ and $L_\alpha$,
            and the magnetic one $L_M$,
     the overall particle contribution $L_P=L_p+L_\alpha$ is also given.
Note that $-L_\alpha$ instead of $L_\alpha$ is plotted in Fig.\ref{fig_angmom_fast_par}d.
Moreover, the horizontal bars on the left of Figs.\ref{fig_angmom_fast_par}b and \ref{fig_angmom_fast_par}c represent
        the azimuthal ion speeds derived in the isotropic
	case at the corresponding locations for comparison.
The open circles correspond to the cases where $\Gamma_{\alpha E}=1.3$.
It turns out that at any given $\Gamma_{pE}$ each parameter varies monotonically from
      the value with $\Gamma_{\alpha E}=0.7$, represented by the end of 
      the arrow, to the value with $\Gamma_{\alpha E}=1.9$ given by the arrow head.
In Fig.\ref{fig_angmom_fast_par}b the arrows have been slightly shifted from one another to avoid overlapping.

From Fig.\ref{fig_angmom_fast_par}a one can see that $\lambda$ decreases with increasing $\Gamma_{pE}$ or $\Gamma_{\alpha E}$, 
     ranging from $1.101$ at the upper left to $1.058$ at the lower right corner.
The deviation of $\lambda$ from unity, albeit modest, indicates that the changes introduced in the total angular
     momentum loss due to the ion pressure anisotropies are not negligible.
From Figs.\ref{fig_angmom_fast_par}b and \ref{fig_angmom_fast_par}c one can see that
     between 0.3 to 1~AU,
     the magnitude of the azimuthal speeds of both species decreases with increasing distance.
Furthermore, at either 0.3 or 1~AU, both $v_{p\phi}$ and $v_{\alpha\phi}$ increase when
     $\Gamma_{pE}$ or $\Gamma_{\alpha E}$ increases.
Take the values at 1~AU for instance.
One can see that $v_{p\phi}$ ranges from $2.36$ to $3.46$\velunits, while $v_{\alpha\phi}$ 
     varies between $-12.8$ and $-11.7$\velunits.
For the majority of the solutions both $v_{p\phi}$ and $v_{\alpha\phi}$ tend to be larger in the algebraic sense
     than the corresponding values in the isotropic model, which yield 
     $2.54$ and $-12.6$\velunits\ for protons and alpha particles, respectively. 
However, at 0.3~AU $v_{p\phi}$ or $v_{\alpha\phi}$ tends to be smaller in the anisotropic than in the isotropic case.
Now $v_{p\phi}$ and $v_{\alpha\phi}$ vary in the intervals $[2.37, 3.77]$ and $[-15.8, -14.2]$\velunits, respectively.
For comparison, the isotropic model yields a $v_{p\phi}$ ($v_{\alpha\phi}$) of $3.33$ ($-14.7$)\velunits.
Now let us examine the specific angular momentum fluxes $L_p$, $L_\alpha$ and $L_M$ at 1~AU.
Figure~\ref{fig_angmom_fast_par}d indicates that $L_M$ has the weakest
     parameter dependence, which is easily understandable given that to a good approximation $\tan\Phi \approx -A_\Omega R/v_{kl}$ 
     where $k$ may be taken to be $p$ or $\alpha$ (see Eq.(\ref{eq_align_all})).
Besides, the parameter dependence of $L_\alpha$ is
     rather modest, varying by $\lesssim 10$\% from $-3.55$ to $-3.23$ when $\Gamma_{pE}$ or $\Gamma_{\alpha E}$ changes.
On the other hand, $L_p$ changes substantially, ranging between $3.46$ and $5.07$.
Hence the overall particle contribution $L_P$ also shows a significant parameter dependence.
In particular, $L_P$ may exceed $L_M$ when $\Gamma_{pE}\gtrsim 1.3$.
For the solutions examined,
     $L_P$ can be found to be positive and attain its maximum of $L_P=1.84$ when $[\Gamma_{pE}, \Gamma_{\alpha E}]=[1.5, 1.9]$.
Only for the lowest values of $\Gamma_{pE}$ and $\Gamma_{\alpha E}$ can one find a negative $L_P$ of
     $-0.084$.
Moreover, the protons always show a partial corotation, i.e., $L_p >0$.
From this we conclude that the ion temperature anisotropies are unlikely the cause of the tendency for $L_p$ or $L_P$ to be 
     negative for the fast solar wind as indicated by the Helios measurements~\citep{Pizzo_etal_83, MarschRichter_84a}.

\subsection{Slow solar wind}
\label{sec_numres_slow}

\begin{figure}
 \centering
  \includegraphics[width=.49\textwidth]{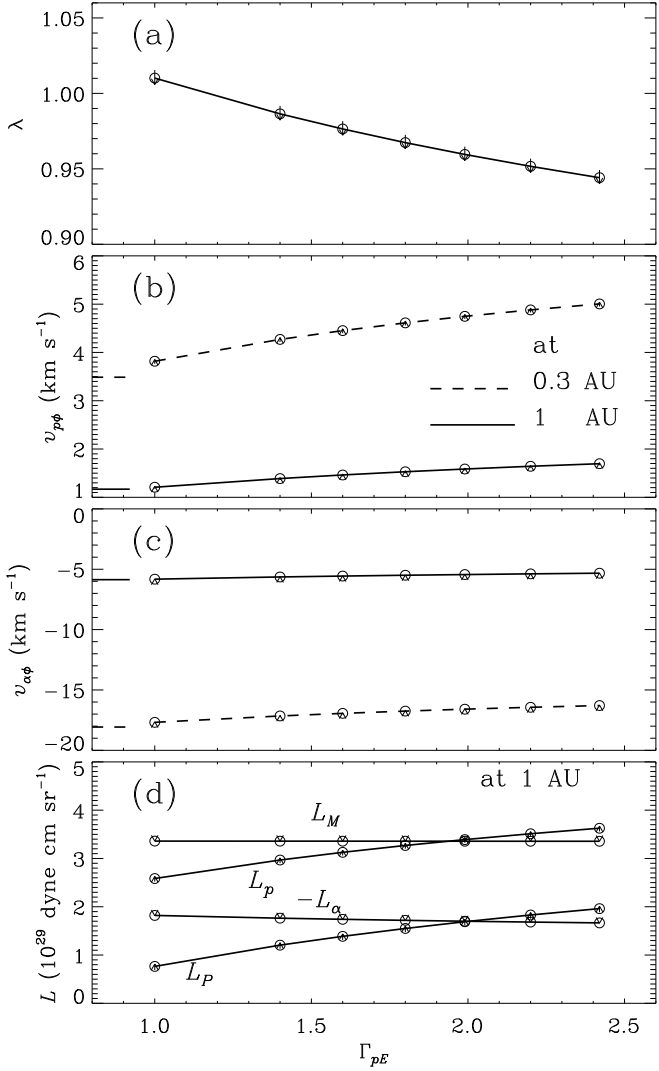}
 \vskip 0.5cm
 \caption{
Similar to Figure~\ref{fig_angmom_fast_par} but for the slow solar wind.
Here the open circles correspond to the cases where $\Gamma_{\alpha E}$ is fixed at $1.4$, and
      the arrow represents how the specific parameter varies at a given $\Gamma_{p E}$
      when $\Gamma_{\alpha E}$ increases from $0.8$ to $2.0$.
}
\label{fig_angmom_slow_par}
\end{figure}

Figure~\ref{fig_angmom_slow_par} presents, in the same fashion as Fig.\ref{fig_angmom_fast_par}, the dependence on 
      $\Gamma_{pE}$ and $\Gamma_{\alpha E}$ of various quantities derived for the
      slow solar wind.
A comparison with Fig.\ref{fig_angmom_fast_par} indicates that nearly all the features in Fig.\ref{fig_angmom_slow_par}
      are reminiscent of those obtained for fast solar wind solutions.
However, some quantitative differences exist nonetheless.
For instance, when $\Gamma_{pE}$ is held fixed, all the examined parameters for the slow wind vary little even though $\Gamma_{\alpha E}$
      changes considerably from $0.8$ to $2$.
In contrast, the parameters for the fast wind show an obvious $\Gamma_{\alpha E}$ dependence.
This difference can be largely attributed to the fact that in the slow wind the ions are
      substantially cooler than in the fast wind.
Figure~\ref{fig_angmom_slow_par}a shows that $\lambda$ ranges from $0.94$ to $1.016$.
In other words, relative to the isotropic case, the solar angular momentum loss rate per steradian
      in the anisotropic models may be enhanced or reduced by up to $6\%$.
If examining Figs.\ref{fig_angmom_slow_par}b and \ref{fig_angmom_slow_par}c, one may find that at both $0.3$ and $1$~AU,
      the azimuthal speeds of both ion species, $v_{p\phi}$ and $v_{\alpha\phi}$, are
      larger algebraically in the anisotropic models than in the isotropic one.
The difference between the two is more prominent at 0.3 AU, where the isotropic model yields that 
      $[v_{p\phi}, v_{\alpha\phi}]=[3.49, -18.1]$\velunits, whereas the anisotropic models yield
      that with increasing $\Gamma_{pE}$, $v_{p\phi}$ increases 
      from $3.76$ to $5.04$\velunits, and $v_{\alpha\phi}$ varies between $-17.8$ to $-16.3$\velunits. 
As for the ion azimuthal speeds at 1~AU, one can see that varying $\Gamma_{pE}$ leads to a $v_{p\phi}$ varying between $1.18$
      and $1.72$\velunits, and a $v_{\alpha\phi}$ ranging from $-5.85$ to $-5.31$\velunits. 
The corresponding changes in the specific ion angular momentum fluxes are shown by Fig.\ref{fig_angmom_slow_par}d,
      which indicates that the proton one $L_p$ increases with increasing $\Gamma_{pE}$ from
      $2.52$ to $3.67$, and likewise, the alpha one $L_\alpha$ increases from $-1.83$ to $-1.66$.
On the other hand, the flux associated with magnetic stresses $L_M$ hardly varies, and a value of $3.36$ can be quoted for all the models examined.
Therefore in the parameter space explored, $L_M$ may be smaller than $L_p$, which is however offset by the alpha contribution
      that is always in the direction of counter-rotation to the Sun.
In fact, the alpha contribution is so significant that the overall particle contribution $L_P$ never exceeds $L_M$. 
In other words, incorporating ion temperature anisotropy cannot resolve the outstanding discrepancy between previous models and
      observations concerning the relative importance of particle and magnetic contributions in the angular momentum budget of
      the solar wind.

\section{Discussion}
\label{sec_discussion}
As demonstrated by~\citet{LiLi08}, the discussion on the angular momentum transport also allows us to say
     a few words on the frequency spectra $S_k (f)$ ($k=p, \alpha$) of the ion velocity fluctuations 
     during Alfv\'enic activities in the fast solar wind in
     the super-Alfv\'enic portion where $M_T^2 \gg 1$.
This is due to the well-known change of the properties of Alfv\'enic fluctuations around some
     $f_c \approx v_{cm, a}/(4\pi r_a)$, where $v_{cm, a}$ is the speed of center of mass evaluated
     at the Alfv\'en point $r_a$~\citep[see e.g., ][]{HO80, LiLi08}.
For typical fast wind parameters, $f_c \approx 0.5-1 \times 10^{-5}$\frequnits.
While the fluctuations with frequencies $f\gtrsim f_c$ are genuinely wave-like
     and may be described by the WKB limit given
     the slow spatial variation of flow parameters in the region in question,
     those with $f\lesssim f_c$ behave in a quasi-static manner and may be described by the solutions to the angular momentum
     conservation law which also governs the zero-frequency fluctuations.
As shown by \citet{LiLi08} who neglected the species temperature anisotropy,
     in the region $r\gtrsim 0.2$~AU which will be explored by the Solar Orbiter and Solar Probe,
     the ratio of the alpha to proton velocity fluctuation amplitude $\delta u_\alpha/\delta u_p$
     can be an order-of-magnitude larger for $f<f_c$ than for $f>f_c$.
Hence one may expect that, if the proton velocity fluctuation spectrum $S_p (f)$ is somehow smooth around $f_c$,
     then the alpha one $S_\alpha (f)$ will show an apparent spectral break.
Now let us revisit this problem in light of the discussion presented in this paper and see what changes the pressure
     anisotropies may introduce.

\begin{figure}
 \centering
 \includegraphics[width=.45\textwidth]{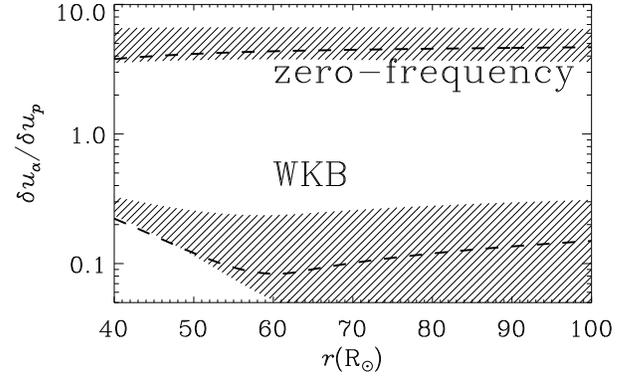}
 \vskip 0.5cm
 \caption{
Radial dependence of the ratio of the alpha to the proton velocity fluctuation amplitudes $\delta u_\alpha/\delta u_p$
    induced by Alfv\'enic activities in super-Alfv\'enic portions of the fast solar wind.
The dashed curves correspond to the isotropic model, while the hatched areas give the possible range $\delta u_\alpha/\delta u_p$
    may occupy when the parameters $\Gamma_{pE}$ and $\Gamma_{\alpha E}$ vary in the ranges given in text.
Both the zero-frequency (upper portion) and WKB (lower) estimates are given.
}
\label{fig_discuss_spectra}
\end{figure}

Restrict ourselves to either the high-latitude region or the region inside say $100$~$R_\odot$ such that the
     magnetic field may be seen as radial.
Furthermore, suppose that the waves are propagating parallel to the magnetic field in the empirical
     fast wind profiles detailed in section~\ref{sec_presflow}.
Figure~\ref{fig_discuss_spectra} presents the radial dependence of $\delta u_\alpha/\delta u_p$ in the region between $40$
     and $100$~$R_\odot$ for both the zero-frequency (upper part) and WKB (lower part) solutions.
For comparison, the dashed curves represent the corresponding results in the isotropic model.
To construct Fig.\ref{fig_discuss_spectra}, all the possible values of $\Gamma_{pE}$ and $\Gamma_{\alpha E}$ have been examined.
As a result, at any radial location the ratio $\delta u_\alpha/\delta u_p$ varies from model to model, and the range in which
     this ratio may occupy is given by the hatched area.
The zero-frequency solutions are obtained by solving Eq.(\ref{eq_tanp_standard}), while for hydromagnetic
     WKB Alfv\'en waves it is well known that $\delta u_\alpha /\delta u_p = |(v_{ph}-v_{\alpha})/(v_{ph}-v_{p})|$,
     where $v_{ph}$ is the wave phase speed and given by \citep[e.g.,][]{BarnesSuffolk_71, I84}
\begin{eqnarray*}
 v_{ph} = v_{cm} + \sqrt{v_{A}^2\left(1-\frac{4\pi P^\Delta}{B^2}\right)-\hat{\rho}_\alpha \hat{\rho}_p v_{\alpha p}^2} \hspace*{0.5cm},
\label{eq_phase_alfven}
\end{eqnarray*}
       in which $v_{cm} = \hat{\rho}_p v_{p} + \hat{\rho}_\alpha v_{\alpha}$ is the speed of center of mass,
       and $\hat{\rho}_k = \rho_k/\rho$ ($k=p, \alpha$) defines the fractional ion mass density.

From Fig.\ref{fig_discuss_spectra} one can see that the zero-frequency and WKB solutions are well separated from each other,
       in the isotropic and anisotropic cases alike.
For the isotropic model, $\delta u_\alpha/\delta u_p$ in the zero-frequency case increases monotonically from
      $3.79$ at 40~$R_\odot$ to 4.68 at 100~$R_\odot$.
On the other hand, in the WKB case it decreases first from 0.22 at 40~$R_\odot$ and attains its minimum of $0.083$ at $60.3$~$R_\odot$
      and then increases to 0.15 at 100~$R_\odot$.
The difference in $\delta u_\alpha/\delta u_p$ between the zero-frequency and WKB solutions may be slightly smaller in the anisotropic
      than in the isotropic case for some combinations of $[\Gamma_{p E}, \Gamma_{\alpha E}]$,
      but the difference is still quite significant.
From this we can conclude that, with realistic ion temperature anisotropies included, 
      the alpha velocity fluctuation spectrum $S_\alpha (f)$
      during Alfv\'enic activities will also show an apparent break near $f_c$, if the proton one $S_p (f)$ is smooth there.
This break is entirely a linear property, and has nothing to do with the nonlinearities that may also shape the fluctuation spectra.

\section{Summary}
\label{sec_summary}

This study has been motivated by the apparent lack of an analysis on the angular momentum transport in a multicomponent solar or stellar
      wind with differentially flowing ions and species temperature anisotropy.
Moreover, there has been an outstanding discrepancy between available measurements and models concerning the relative
      importance of the particle $L_P$ and magnetic contribution $L_M$ to the solar angular momentum loss rate per steradian $L$.
The Helios measurements indicate that for fast (slow) solar wind with $v_p\gtrsim 600$ ($\lesssim 400$)\velunits,
      $L_P$ tends to be negative (positive), with the positive sign denoting the direction of corotation with the Sun.
Furthermore, $L_P$ tends to be larger than $L_M$ in the slow wind.
The behavior of $L_P$ derives from that of individual ion angular momentum fluxes, $L_p$ and $L_\alpha$,
      thereby calling for a multifluid approach.

Starting with a general set of multifluid transport equations with gyrotropic species pressure tensors, we have derived the equations 
     for both the angular momentum conservation (Eqs.(\ref{eq_align_all}) and (\ref{eq_tanp_standard}) in section~\ref{sec_mathform}),
      and the energy and linear momentum balance (Eqs.(\ref{eq_reduced_nk}) to (\ref{eq_reduced_Tsper}) in the appendix).
These equations are not restricted to radial outflows in the equatorial plane, instead they are valid for arbitrary axisymmetrical
      winds that include two major ion species, and therefore are expected to find applications in general outflows from late-type stars.
To focus on the problem of angular momentum transport, we refrained from solving the full set of equations governing the meridional
      dynamics.
Rather, we constructed, largely based on the available in situ measurements, the empirical profiles for the meridional magnetic field and flow parameters. 
Only the ion temperature anisotropies are considered, i.e., the electron temperature is seen as isotropic.
For both the fast and slow solar wind profiles, we solved the angular momentum conservation law (Eqs.(\ref{eq_align_all}) and (\ref{eq_tanp_standard})) to examine
      how the azimuthal speeds of protons $v_{p\phi}$ and alpha particles $v_{\alpha\phi}$,
      as well as the individual components in the solar angular momentum budget
      are influenced by the ion temperature anisotropies.
To this end, solutions to the isotropic version are obtained for comparison.

Our main conclusions are:
\begin{enumerate}
 \item 
  From the derived equations governing the energy transport, a simple analysis given in the appendix yields that the adiabatic cooling may be considerably influenced with the introduction of the azimuthal components. 
  Such an influence is understandably more prominent
     in the low-latitude regions. 
  This means, when modeling the species temperature anisotropy, for a quantitative comparison of model computations to be made with the near-ecliptic measurements such as made by Helios, the spiral magnetic field has to be taken into account.
 \item In agreement with the single-fluid case~\citep{WD_70, Weber_70}, incorporating species temperature anisotropy leads
      to a situation where the total angular momentum loss rate per steradian $L$ is determined by the behavior of the solution to the
      angular momentum conservation law in the vicinity 
      of the Alfv\'en point where the combined Alfv\'enic Mach number $M_T=1$. However, $M_T$ has to take into account the
      contribution from both ion species, as defined by Eq.(\ref{eq_def_MT2_betaDelta}).
\item Relative to the isotropic case, the introduced species temperature anisotropy may enhance or decrease $L$ by up to $10$\%,
      and introduce an absolute change of up to $\sim 1.8$\velunits\ in individual ion azimuthal speeds in the region between $0.3$ and $1$~AU.
      While these changes seem modest, the corresponding changes in the angular momentum fluxes convected by protons $L_p$ or alpha particles $L_\alpha$ may change substantially. 
      In contrast, the flux associated with magnetic stresses $L_M$ hardly varies.
\item However, introducing ion temperature anisotropies cannot resolve the discrepancy
      between in situ measurements and models.
For the fast wind solutions, while in extreme cases $L_P$ may become negative $L_p$ always stays positive.
On the other hand, for the slow solar wind solutions examined, $L_P$ never exceeds $L_M$ even though $L_M$ may be smaller than the
     individual ion contribution.
This is because, for both the slow and fast wind solutions, $L_p$ and $L_\alpha$ always have opposite signs.
\item The discussion on the angular momentum transport has some bearing on the ion velocity fluctuation spectra
      $S_k (f)$ ($k=p, \alpha$) during
      Alfv\'enic activities in the super-Alfv\'enic regions, which are likely to be explored by future missions such as Solar Orbiter and Solar Probe.
      In agreement with~\citet{LiLi08} where species temperature anisotropies are neglected, an analysis based on the WKB and zero-frequency solutions yields that $S_\alpha (f)$ will show an apparent break around some critical frequency $f_c$ if $S_p (f)$ is smooth there.
      This $f_c \sim 0.5-1\times 10^{-5}$\frequnits\ is the well-known frequency that separates the genuinely wave-like fluctuations from quasi-static ones. 
\end{enumerate}

\begin{acknowledgements} 
We thank the referee (Dr. Horst Fichtner) for his very helpful comments.
This research is supported by an STFC rolling grant to Aberystwyth University.
\end{acknowledgements}

\Online
\begin{appendix} 
\section{Derivation of equations governing the meridional dynamics}
In section \ref{sec_mathform}, we have demonstrated that the vector equations governing a time-independent
    multicomponent solar wind with species
    temperature anisotropy are allowed to be decomposed into a force balance condition across the poloidal
    magnetic field and a set of transport equations along it. 
The azimuthal dynamics has been discussed in the text, whereas this appendix provides some discussion 
    on the poloidal dynamics.
In particular, we shall derive the equations governing the poloidal motion $v_{k l}$ of ion species ($k=p, \alpha$),
    and the species temperatures $T_s^{\parallel, \perp}$ ($s=e, p, \alpha$)
    in rather general situations.

Due to the presence of $v_{kN}$ in the $l$ component of the ion momentum equation~(\ref{eq_gen_vec_vk}),
    one may expect that the $N$-component of Eq.(\ref{eq_gen_vec_vk}) 
    has to be solved. 
In fact, there is no need to do so because $v_{k N}$ appears only in the difference $v_{j N}-v_{k N}$, which
    may be found from the $\phi$ component of Eq.(\ref{eq_gen_vec_vk}).
Substituting $v_{j N}-v_{k N}$ into the $l$ component of Eq.(\ref{eq_gen_vec_vk}) will then eliminate
   the cumbersome $\Omega_k$ and $v_{k N}$.
Note that this technique, first devised by \citet{McKenzie_etal_79}, ensures
     the conservation of not only total momentum but also total energy
     \citep[see][]{LiLi06}.
Specifically, the resulting equations for the poloidal dynamics are
\begin{eqnarray}
&& \left(\frac{n_k v_{k l}}{B_l}\right)' = 0, \label{eq_reduced_nk} \\
&& v_{k l}(v_{k l})' - v_{k \phi}^2 (\ln R)' + \tan\Phi \frac{v_{k l}}{R}(R v_{k \phi})' \nonumber \\
    &-&   (C_{k l} + \tan\Phi C_{k \phi}) + \frac{G M_\odot}{r}(\ln r)' \nonumber \\
    &+& \frac{1}{n_k m_k}\left\{(p_k^\parallel)'-p_k^\Delta\left[\ln \left(B_l\sec\Phi\right)\right]'\right\} \nonumber\\
    &+& \frac{Z_k}{n_e m_k}\left\{(p_e^\parallel)'-p_e^\Delta\left[\ln \left(B_l\sec\Phi\right)\right]'\right\} = 0, \label{eq_reduced_vkl} \\
&& v_{s l} \left(T_s^{\parallel}\right)' 
    + 2 v_{s l} T_s^{\parallel} \left[\ln\left(v_{s l}\sec\Phi\right)\right]' \nonumber \\
   &+&  \frac{1}{n_s k_B}\left[\nabla\cdot \vec{q}_s^\parallel  
         -\tens{Q}_s\vdots\nabla(\hat{b}\hat{b})-\frac{\delta E_s^\parallel}{\delta t} - H_s^\parallel\right] =0, \label{eq_reduced_Tspara}\\
&& v_{s l} \left(T_s^{\perp}\right)' 
    - v_{s l}T_s^{\perp}\left[\ln\left(B_{l}\sec\Phi\right)\right]'  \nonumber \\
   &+&  \frac{1}{n_s k_B}\left[\nabla\cdot \vec{q}_s^\perp  
         +\frac{\tens{Q}_s}{2}\vdots\nabla(\hat{b}\hat{b})-\frac{\delta E_s^\perp}{\delta t} - H_s^\perp\right] =0, \label{eq_reduced_Tsper}
\end{eqnarray}
     where $p_s^\Delta = p_s^\parallel-p_s^\perp$ ($s=e, p, \alpha$), $\vec{C}_k = \delta \vec{M}_k/\delta t + (Z_k n_k/n_e)\delta \vec{M}_e/\delta t$ results from
     Coulomb frictions, and the prime $'$ represents the derivative with respect to the arclength $l$.
When deriving Eqs.(\ref{eq_reduced_Tspara}) and (\ref{eq_reduced_Tsper}), we have used
     the fact that the expression
\begin{eqnarray*}
 \nabla_\parallel\cdot\vec{v}_s 
    &=& \cos^2\Phi v_{sl}' + \sin^2\Phi v_{sl} (\ln R)' \\
    &+& \cos\Phi\sin\Phi R(v_{s\phi}/R)',
\end{eqnarray*}
     may be simplified by expressing $v_{s\phi}$ via the alignment condition~(\ref{eq_align_all}), 
     the result being
\begin{eqnarray*}
 \nabla_\parallel\cdot\vec{v}_s 
   = v_{sl}\left[\ln \left(v_{sl}\sec\Phi\right)\right]' .
\end{eqnarray*}
Similarly, one may find that
\begin{eqnarray*}
 \nabla_\perp\cdot\vec{v}_s 
   = -v_{sl}\left[\ln \left(B_{l}\sec\Phi\right)\right]' .
\end{eqnarray*}

Now let us compare these equations with those in \citet{I84}. 
It is straightforward to show that Eqs.(\ref{eq_reduced_Tspara}) and (\ref{eq_reduced_Tsper}) 
      are equivalent to Eqs.(A3) and (A4) in \citet{I84} by specializing to a spherically
      symmetric solar wind and by noting that $(\ln B_l)' = -2/r$.
On the other hand, using the alignment condition (\ref{eq_align_all}) one may find that 
\begin{eqnarray}
&& v_{k l} \left(v_{k l}\right)' - v_{k \phi}^2 \left(\ln R\right)'
       + \tan\Phi \frac{v_{k l}}{R}\left(R v_{k \phi}\right)' \nonumber \\
&=& \left(\frac{v_{k l}^2}{2} \sec^2\Phi\right)'
  -\left(\frac{A_\Omega^2 R^2}{2}\right)' . \label{eq_vk_corot}
\end{eqnarray} 
Again specializing to a spherical solar wind, one then finds that Eq.(\ref{eq_reduced_vkl}) is equivalent to 
      (A2) in \citet{I84}.
It should be stressed that although working in a frame of reference corotating with the Sun, as did \citet{I84},
      substantially simplifies the algebra, it does not offer the information on the specific form
      of the spiral angle $\Phi$, whose functional dependence on the flow speeds has to be assumed a priori.
In practice, \citet{I84} assumed that the velocity of center of mass $\vec{v}_{cm}$ is radial in an inertial frame
      beyond 10~$R_\odot$, which is certainly a good assumption for the present slow-rotating Sun.
However, from our discussion on the azimuthal dynamics, there is in general no guarantee that
      $\vec{v}_{cm}$ is radial, and the deviation may be substantial for 
      winds that flow from a faster rotating star.

Introducing azimuthal components may influence the ion flow speeds $v_{k l}$ both directly and indirectly.
The direct consequence is that azimuthal components may introduce into the reduced meridional momentum
      equation~(\ref{eq_reduced_vkl})
      an effective force (see the first three terms). 
Note that in a corotating frame the magnitude of the ion velocity becomes $v_{k l}\sec\Phi$, from 
     relation~(\ref{eq_vk_corot}) one may see that in such a frame all particles
     move in the same centrifugal potential $A_\Omega^2 R^2/2$.
Therefore in effect the introduced force tends to reduce the magnitude of the ion speed difference
     with increasing distance as $\sec\Phi$ tends to increase. 
This effect has been explored in detail in \citet{LiLi06} and \citet{Li_etal_07},
     where it is shown that the influence may play an important part in the force balance for the solar wind.
In fact, introducing solar rotation alone is able to reproduce the $v_{\alpha p}$ profile measured by Ulysses
     beyond 2~AU if a proper value of $v_{\alpha p}$ is imposed there.
On the other hand, $v_{k l}$ may be altered indirectly by the modified pressure
     gradient force due to changes in the temperatures, which in turn are caused by the
     changes in the heat fluxes (the third term in Eqs.(\ref{eq_reduced_Tspara})
     and (\ref{eq_reduced_Tsper})) and through the adiabatic cooling (the second term).
A detailed discussion on the former requires a specific form for the heat flux, which is beyond the
     scope of the present paper.
As a consequence, we shall focus on the latter instead.

Neglecting the terms in the second pair of square parentheses, Eqs.(\ref{eq_reduced_Tspara})
     and (\ref{eq_reduced_Tsper}) give 
\begin{eqnarray}
 T_s^{\parallel} \propto \cos^2\Phi/v_{s l}^2, T_s^{\perp} \propto B_l\sec\Phi.
\end{eqnarray}
Note that the relation governing $T_s^{\perp}$ simply reflects the conservation of magnetic moment.
Now that in the region say $r>10$~$R_\odot$ $\sec\Phi$ is significant and increases with $r$,
     $T_s^{\parallel}$ ($T_s^{\perp}$) may be substantially reduced (enhanced) relative to the case
     where $\Phi\equiv 0$.
This effect is particularly significant in the near-ecliptic region and for the slow solar wind.
For instance, restrict ourselves to the equatorial plane and consider the region between say 10~$R_\odot$ and 1~AU.
Suppose $v_{s l}$ remains constant and $v_{s l} \approx A_\Omega R_E = 430$~km/s.
Now that roughly speaking $\tan\Phi \approx -A_\Omega r/v_{s l}$, 
    when the spiral field is considered, $T_s^{\parallel}$ ($T_s^\perp$) at 1~AU is $1/2$ ($\sqrt{2}$) times the value for a purely radial magnetic field.
This suggests that for making any quantitative comparison of the modeled species
    temperature anisotropy with the near-ecliptic measurements
    such as made by Helios, the spiral magnetic field has to be considered.

For completeness, we note that the force balance condition across the $N$ direction comes from 
    the $N$ component of the total momentum, which reads
\begin{eqnarray}
&&  \sum_k \rho_k \left( \frac{v_{k l}^2}{R_c} - v_{k \phi}^2 \frac{\partial}{\partial N} \ln R\right) 
 + \frac{\partial}{\partial N}\left(P^{\perp} + \frac{B^2}{8\pi}\right) \nonumber \\
&-& \frac{1}{4\pi}\left(1-\frac{4\pi P^\Delta}{B^2}\right)
       \left(\frac{B_l^2}{R_c} - B_{\phi}^2 \frac{\partial}{\partial N} \ln R\right) \nonumber \\
 &+& \sum_k \rho_k \frac{G M_\odot}{r}\frac{\partial}{\partial N} \ln r = 0,
\end{eqnarray}
      where $R_c = \hat{e}_N \cdot \left(\hat{e}_l\cdot \nabla\hat{e}_l\right)$ is the signed curvature radius of the poloidal magnetic line of force.
Obviously, this force balance condition 
      determines the poloidal magnetic field configuration
      in response to the electric currents associated with the flow.
This equation, combined with the transport equations along the meridional magnetic line of force,
      may be solved alternately
      to find a self-consistent solution to the vector equations
      by using the approach by \citet{PneumanKopp_71} or \citet{Sakurai_85}.
\end{appendix}

\end{document}